\documentclass[12pt]{iopart}

\usepackage[dvips]{graphicx}
\begin{document}
\title{The ordering of {\it XY} spin glasses
}
\author{Hikaru Kawamura}
\address{Department of Earth and Space Science, Faculty of Science,
Osaka University, Toyonaka 560-0043,
Japan}
\ead{kawamura@ess.sci.osaka-u.ac.jp}
\date{\today}
\begin{abstract}
Ordering properties of {\it XY\/}-like spin-glass magnets with an easy-plane magnetic anisotropy are studied based on  a symmetry consideration and the results of recent numerical simulations on the pure Heisenberg and  {\it XY\/} spin-glass models. The effects of an easy-plane-type uniaxial anisotropy, a random magnetic anisotropy and an applied magnetic field are investigated. In the {\it XY\/} regime in zero field, the ``spin-chirality decoupling'' persists even under the random magnetic anisotropy, escaping the ``spin-chirality recoupling'' phenomenon which inevitably occurs in the Heisenberg regime. Contrast between the scalar chiral order and the vector chiral order is emphasized. Implications to experiments are discussed.
\end{abstract}
%
%
%
%\maketitle
%
%
%

\section{Introduction}

Spin-glass (SG) magnets are the type of random magnets in which both ferromagnetic and antiferromagnetic interactions coexist and compete, thereby giving rise to the effects of frustration and quenched randomness. Most typical examples might be the so-called  ``canonical SG'',  dilute transition-metal alloys soluted in the noble-metal host such as AuFe and CuMn \cite{review}.  Experimentally, we now have fairly convincing evidence that typical SG magnets exhibit an equilibrium  phase transition at a finite temperature into the glassy ordered state. The true nature of the SG ordering, however, still remains elusive and has been hotly debated \cite{review}. 

 It should be noticed that most of real SG magnets including canonical SGs are of the Heisenberg-type, {\it i.e.\/}, the magnetic anisotropy, which on average possesses no preferential direction in spin space, is considerably weaker in magnitude than the main isotropic exchange interaction. For the fully isotropic  3D Heisenberg SG, it was proposed in as early as 1992 that the model might exhibit an intriguing ``spin-chirality decoupling'' phenomenon \cite{Kawamura92}. Namely, the chirality might be ``decoupled'' from the spin in the sense that the chiral-glass (CG) order takes place at a temperature higher than the SG order, $T_{CG} > T_{SG}$ \cite{Kawamura98,KawaIma,ImaKawa,HukuKawa00,HukuKawa05,VietKawamura09}. Based on such a spin-chirality decoupling picture of the 3D isotropic Heisenberg SG, a chirality scenario of experimental SG transition was advanced \cite{Kawamura92,Kawamura07,Kawamura10}: According to this scenario, the chirality is a hidden order parameter of real SG transition. Real SG transition of weakly anisotropic SG magnets is then a ``disguised'' CG transition, where the chirality is mixed into the spin sector via a weak random magnetic anisotropy. Although the validity of the spin-chirality decoupling picture and the chirality scenario of experimental SG ordering has long been under debate \cite{LeeYoung03,Campos06,Campbell07,LeeYoung07,Fernandez}, recent simulations seem to give some support to the occurrence of such a spin-chirality decoupling in the 3D Heisenberg SG \cite{VietKawamura09} and related models \cite{VietKawamura1D}. 

 Chirality in the Heisenberg SG is the so-called ``scalar chirality'', which is defined for three neighboring spins by the scalar $\chi =\vec S_1\cdot (\vec S_2\times \vec S_3)$. This quantity takes a nonzero value for any noncoplanar spin configuration while it vanishes identically for any coplanar spin configuration. The sign of the scalar chirality indicates whether the noncoplanar spin configuration is either right- or left-handed, labelling the $Z_2$ chiral degeneracy inherent to the frustration-induced noncoplanar spin configuration. Note that the scalar chirality changes its sign under time-reversal operation $\vec S_i \rightarrow -\vec S_i$ while it remains invariant under $SO(3)$ `spin-rotation'  in spin space, $\vec S_i \rightarrow R\vec S_i$ ($R\in SO(3)$). A word of caution here: In the present paper, we use a terminology `spin-rotation'  as representing a spin-rotation {\it performed solely in spin space\/}, without accompanying rotation in real space. Note that such an `internal' symmetry operation, though still remain to be a well-defined symmetry operation, often differs from the standard rotation operation which accompanies real-space rotation, and we use here the term with quotation marks. After all, in random systems like SGs, the standard real-space rotational symmetry is absent. Similar convention will be understood below also for `spin-reflection' operation.

 Other type of chirality, the so-called ``vector chirality'', has also been known \cite{Villain}.  The vector chirality is defined for two neighboring spins by the vector $\vec \kappa =\vec S_1\times \vec S_2$ \cite{KawaTane85,KawaTane87}. (While the vector chirality is often defined on a plaquette consisting of four bonds in numerical simulations \cite{KawaTane85,KawaTane87}, such difference in its definition is irrelevant to the essential features of underlying physics.) It becomes  nonzero for any noncollinear spin configuration, even for coplanar spin configuration,  but vanishes identically for any collinear spin configuration. 

 In the case of two-component {\it XY\/} spins $\vec S=(S_{x}, S_{y})$ ordered in a noncollinear manner, the vector chirality has only a $z$-component, the sign of which tells whether the noncollinear spin configuration in the {\it XY\/}-plane is either right- or left-handed. Thus, for two-component {\it XY\/} spin systems including the {\it XY\/} SG, the $z$-component of the vector chirality $\kappa_z$ represents a discrete $Z_2$ chiral degeneracy \cite{Villain,KawaTane85,KawaTane87}, just as the scalar chirality $\chi$  represents a discrete $Z_2$ chiral degeneracy for three-component Heisenberg spin systems.  Meanwhile, the vector chirality $\kappa_z$ of the {\it XY\/} spin exhibits different symmetry properties from the scalar chirality $\chi$ of the Heisenberg spin. Namely, $\kappa_z$ remains invariant under time-reversal operation, but changes its sign under global internal `spin-reflection' operation in two-component spin space to be defined below. 

 In the present paper, we are interested in the ordering properties of {\it XY\/}-like SG magnets, {\it i.e.\/}, the vector SG with an easy-plane-type magnetic anisotropy. Such a system possesses a nontrivial $Z_2$ chiral degree of freedom associated with the vector chirality. For the two-component {\it XY\/} SG in three spatial dimensions (3D), the possibility of the spin-chirality decoupling has also been suggested, {\it i.e.\/}, successive CG and SG transitions occurring at $T=T_{CG}$ and $T=T_{SG}$ with $T_{CG} > T_{SG}$ \cite{KawaTane91}. Some numerical support for such a decoupling was reported \cite{KawaXY92,KawaXY95,KawaLiXY}, but some others claimed a simultaneous occurrence of spin and chiral transitions \cite{Granato3D,Nakamura04,PixleyYoung08}. Hence, the subject still remains somewhat controversial. 

 Though the number of {\it XY\/}-like SG materials is rather limited compared with the Heisenberg-like ones, recent experimental studies have provided several promising candidate materials.  Experimental realization of the {\it XY\/} SG might be found in certain SG magnets with an easy-plane-type uniaxial anisotropy, {\it e.g.\/}, Rb$_2$Mn$_{1-x}$Cr$_x$Cl$_4$ \cite{Katsumata}, CdMn \cite{Murayama}, Eu$_{0.5}$Sr$_{1.5}$MnO$_4$ \cite{Mathieu} and Mn$_x$Ni$_{1-x}$TiO$_3$ \cite{Ito,Kimura}. Other interesting experimental realization might be granular cuprate superconductors consisting of random Josephson network of sub-micron-size superconducting grains \cite{Matsuura,Kawa-orbital,KawaLi,Uppsala}. In the present paper, we shall focus on the first category, {\it i.e.\/}, easy-plane-type SG magnets.  Readers interested in the second category, {\it i.e.\/}, granular cuprate superconductors, might refer to the references \cite{Kawamura10,Matsuura,Kawa-orbital,KawaLi,Uppsala}.

 Interest in the {\it XY\/}-like SG is also promoted by the recent progress in the study of multiferroic properties of certain frustrated magnets which are induced by the vector chirality. It now becomes increasingly clear that the ferroelectric polarization often arises from the vector chiral order realized in frustrated spin systems via the multiferroic coupling (the spin-current mechanism) \cite{Katsura}. Yet, the experimental study on the {\it XY\/}-like SG still remains as a primitive stage. Few systematic studies on the effect of magnetic fields and magnetic anisotropy were done so far. Numerical studies done in the past were also limited to the simplest case of a fully isotropic system in  zero field without a random magnetic anisotropy \cite{KawaTane85,KawaTane87,KawaTane91,KawaXY92,KawaXY95,KawaLiXY,Granato3D,Nakamura04,PixleyYoung08}. It is thus highly desirable to examine the effects of applied magnetic fields, uniaxial and random magnetic anisotropies in a systematic manner.

 Under such circumstances, the aim of the present paper is to explore the ordering properties of {\it XY\/}-like SG magnets with an easy-plane-type magnetic anisotropy on the basis of the recently developed spin-chirality decoupling picture and to promote further experimental studies on {\it XY\/}-like SG magnets. We study the effects of an uniaxial magnetic anisotropy, a random magnetic anisotropy and an applied magnetic field from the standpoint of the chirality picture. These effects are crucially important in properly interpreting and understanding the relevant experimental data. Contrast is then made to the properties of the Heisenberg-like SG magnets. Recent experimental data are discussed.

\section{Model and symmetries}

 Our analysis is based mainly on a symmetry consideration, and in this sense, expected to be insensitive to microscopic details of each particular system. Nevertheless, to fix the idea, it might be useful to present an explicit model Hamiltonian. We first consider the classical Heisenberg Hamiltonian of the form,
\begin{eqnarray}
{\cal H}=-\sum_{<ij>}J_{ij}\vec S_i\cdot \vec S_j + D\sum_i (S_i^z)^2 - H \sum_i S_i^z,
\end{eqnarray}
or of the form,
\begin{eqnarray}
{\cal H}=-\sum_{<ij>}J_{ij} \{ (1+A)(S_i^x S_j^x + S_i^y S_j^y) + S_i^z S_j^z \} 
\ - H \sum_i S_i^z,
\end{eqnarray}
where $\vec S_i$ denotes a three-component classical Heisenberg spin of unit length ($|\vec S_i|=1$) at the site $i$,  $J_{ij}$ is an isotropic random exchange coupling between the sites $i$ and $j$ taking both positive and negative values, $H$ is a magnetic-field intensity, and the sum $\langle ij\rangle$ is taken over spin pairs at the sites $i$ and $j$. The quantities  $A>0$ and $D>0$ represent the strength of an easy-plane-type uniaxial magnetic anisotropy either  of a single-ion-type ($D$) or of an exchange-type ($A$), $D=0$ or $A=0$ corresponding to the isotropic Heisenberg case.

  Reflecting the random local environment of each magnetic ion, real {\it XY\/} SG materials should possess in addition a random magnetic anisotropy which varies from site to site. This random magnetic anisotropy $H^{(R)}$ may be either of a single-ion-type or of an exchange-type as given by
\begin{equation} 
H^{(R)}=\sum_{i}\sum_{\mu,\nu} D^{(R)}_{i,\mu \nu} S_i^{\mu}S_i^{\nu},
\end{equation} 
\begin{equation} 
H^{(R)}=\sum_{\langle ij \rangle}\sum_{\mu,\nu} A^{(R)}_{ij,\mu \nu}S_i^{\mu}S_j^{\nu},
\end{equation} 
where $\mu$ and $\nu$ ($\mu,\nu=x,y,z$) refer to the spin component. In fact, symmetry properties would be common between (1) and (2), or between (3) and (4), and we are not distinguishing (1) and (2), or (3) and (4), in our following discussion.

 We begin with the simplest case of a system in zero field without the random magnetic anisotropy, {\it i.e.\/}, the system described by the Hamiltonian (1) or (2) with $H=0$. Notice our present convention of the term `spin-rotation' or `spin-reflection' mentioned in section 1 above. Furthermore, we mean here by 'spin reflection'  a mirroring operation in spin space {\it performed not taking account of the axial-vector character of the spin\/}. For example, `spin-reflection' with respect to an {\it xy\/}-plane means here an operation $(S_x, S_y, S_z) \rightarrow (S_x, S_y, -S_z)$, which is different from the standard spin-reflection operation taking account of the axial-vector character of the spin, $(S_x, S_y, S_z) \rightarrow (-S_x, -S_y, S_z)$. With these conventions, the Hamiltonians (1) and (2) with $H=0$ possess an invariance under the following global internal symmetry operations.

\medskip\par\noindent
(i) T: $Z_2$ time-reversal operation, $\vec S_i \rightarrow -\vec S_i$.
\par\noindent
(ii) RT$_T$: SO(2) internal `spin-rotation' around the hard axis ($\hat z$), $\vec S_i \rightarrow R\vec S_i$ where $R\in$ SO(2).
\par\noindent
(iii) RT$_L^{(\pi)}$: $Z_2$ internal `spin-rotation' of $\pi$ around an arbitrary axis contained in an $xy$-plane, {\it e.g.\/}, $(S_i^x, S_i^y, S_i^z) \rightarrow (S_i^x, -S_i^y, -S_i^z)$ and similar operations.
\par\noindent
(iv) RF$_T$: $Z_2$ internal `spin-reflection' with respect to an arbitrary plane containing the $z$-axis, {\it e.g.\/}, $(S_i^x, S_i^y, S_i^z) \rightarrow (S_i^x, -S_i^y, S_i^z)$ and other similar operations.
\par\noindent
(v) RF$_L$: $Z_2$ internal `spin-reflection' with respect to an $xy$-plane, $(S_i^x, S_i^y, S_i^z) \rightarrow (S_i^x, S_i^y, -S_i^z)$.

\medskip
Not all symmetry operations given above are mutually independent. For example, time-reversal T is a combination of a $\pi$ `spin-rotation' RT$_T$ and a `spin-reflection' RF$_L$, or a combination of a `spin-reflection' RF$_T$ and  a $\pi$ `spin-rotation' RT$_L^{(\pi)}$. Likewise, RT$_L^{(\pi)}$ is a combination of RF$_T$ and RF$_L$, being not independent of other operations.

 Note that the vector chirality $\kappa_z$ is invariant under the operations T and  RT$_T$, but changes its sign under the operations  RF$_T$ and  RT$_L^{(\pi)}$. Meanwhile, the scalar chirality $\chi$ is invariant under the operations  RT$_T$ and  RT$_L^{(\pi)}$, but changes its sign under the operations T,  RF$_T$ and  RF$_L$. Hence, at the vector-CG transition, the symmetries  RF$_T$ and  RT$_L^{(\pi)}$ should be broken, while, at the scalar-CG transition, the symmetries  T,  RF$_T$ and  RF$_L$ should be broken.

In addition to these global Hamiltonian symmetries, the SG system might sometimes exhibit a peculiar type of symmetry breaking called the replica-symmetry breaking (RSB) unrelated to any global symmetry of the system. In particular, Refs.\cite{KawaLiXY} suggested that the CG order of the 3D {\it XY\/} SG might accompany RSB of one-step-type \cite{KawaLiXY}. Similar one-step-type RSB is also observed in the CG ordered state of the 3D Heisenberg SG \cite{HukuKawa00,Kawamura10}. Hence, we also quote, 

\medskip\noindent
(vi) RS: Replica symmetry  unrelated to any global symmetry of the Hamiltonian, which comes from the nontrivial phase-space structure of the ordered state.

\section{Zero-field properties}

  We construct a phase diagram in the uniaxial anisotropy ($D$ or $A$) versus the temperature ($T$) plane. In the isotropic Heisenberg case $D=0$ or $A=0$, recent numerical studies have indicated that the system exhibits two successive transitions, the scalar-CG transition at a higher temperature  $T=T_{CG}$ and the SG transition at a lower temperature $T=T_{SG} (< T_{CG})$, {\it i.e.\/}, the spin-chirality decoupling \cite{VietKawamura09}. At $T=T_{CG}$, global $Z_2$ `spin-reflection' symmetry and time-reversal symmetry are spontaneously broken  together with a replica symmetry, leading to the scalar-chirality freezing. At $T=T_{SG} < T_{CG}$, $SO(3)$ `spin-rotation' symmetry is broken, leading to the Heisenberg-spin freezing. In the Heisenberg case, the spin is ordered into the noncoplanar configuration, and the chirality relevant to the CG order is the scalar chirality. 

 In the opposite {\it XY\/} limit of $D=\infty$ or $A=\infty$,  `spin-reflection' symmetry RF$_T$ is spontaneously broken at the vector-CG transition $T=T_{CG}$ together with replica symmetry RS, leading to the vector-chirality freezing. Below this vector-CG transition $T\leq T_{CG}$, the phase space exhibits a nontrivial ergodicity breaking due to the 1-step-like RSB. At $T=T_{SG} < T_{CG}$, $SO(2)$ `spin-rotation' symmetry RT$_T$ is broken, leading to the {\it XY\/}-spin freezing. Due to the strong easy-plane anisotropy, the spin order in this case is a coplanar one contained in the {\it XY\/} plane where the scalar chirality vanishes identically. The chirality relevant here is the vector chirality $\kappa_z$.  The issue of whether there occurs an additional RSB at the SG transition or not, and what type if any, is a highly nontrivial one and remains as an open question. Since an RSB (or an ergodicity breaking in the phase space) sets in already at $T=T_{CG}$, possible onset of further additional RSB at $T=T_{TSG}$ would have only secondary effect. In the present paper, we donot pursue this issue further.

 When an {\it XY\/}-like anisotropy is moderate, the SG order might evolve in successive way, each transition associated with the transverse ($xy$) or the longitudinal ($z$) spin component, $S_T$ or $S_L$. Such successive SG transitions were indeed observed in real {\it XY\/}-like SG magnets like CdMn \cite{Murayama} or Mn$_x$Ni$_{1-x}$TiO$_3$ \cite{Ito}. In this parameter range, the system exhibits in total three successive transitions with decreasing the temperature, first the vector-CG transition at $T=T_{CG}$, second the transverse SG transition into the ``coplanar'' SG state at $T=T_{TSG}$, and finally  the longitudinal SG transition into the ``noncoplanar'' SG state at $T=T_{LSG}$. The symmetries spontaneously broken at each transition are RF$_T$ and  RT$_L^{(\pi)}$ at $T_{CG}$,  RT$_T$ and T at $T_{TSG}$, and RF$_L$ at $T_{LSG}$.

 Since the spin and the chirality are decoupled in the Heisenberg limit and the magnetic anisotropy is coupled to the spin, not to the chirality, the scalar-CG state of the fully isotropic Heisenberg case would survive a sufficiently weak magnetic anisotropy. Recall that the SG correlation length remains finite at the CG transition point of the Heisenberg SG. Therefore, we expect a small nonzero anisotropy $D$ or $A$ required to suppress the scalar-CG phase.
 
 In view of these observations, a possible phase diagram of the {\it XY\/}-like SG magnet with an easy-plane-type uniaxial magnetic anisotropy is given in Fig.1 in the anisotropy versus the temperature plane. A multicritical point is expected to appear at a certain strength of uniaxial anisotropy, at which the scalar-CG state vanishes. One can think of a phase diagram  more complicated  than the one shown in Fig.1, particularly near the multicritical point. But, we consider here the simplest one.

\begin{figure}[ht]
\begin{center}
\includegraphics[scale=0.37]{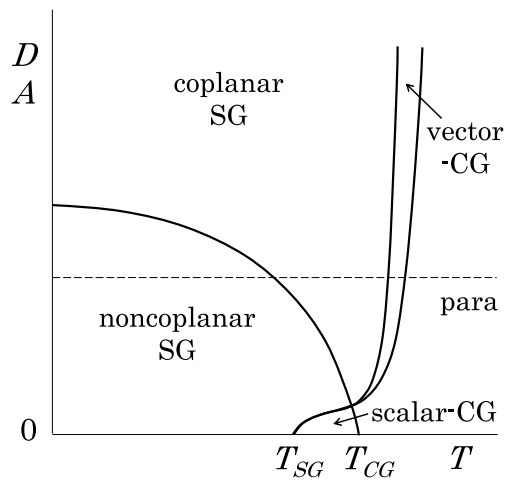}
\end{center}
\caption{
Phase diagram of the {\it XY\/}-like SG with an easy-plane-type uniaxial magnetic anisotropy in the uniaxial anisotropy versus the temperature plane. $T_{CG}$ and $T_{SG}$ represent the chiral-glass and the spin-glass transition temperatures of the fully isotropic Heisenberg system.  
}
\end{figure}

 Ordered states in Fig.1 are characterized by various order parameters such as the scalar chirality $\chi$, the vector chirality $\kappa_z$, the transverse spin component $S_T$ and the longitudinal spin component $S_L$. (In SGs,  as appropriate order parameters, `overlap' variables associated with these quantities and their moments should be considered.)  In Fig.2, the regions of the phase diagram where each order parameter becomes nonzero are indicated by the hatched region.

 Many of real {\it XY\/}-like SG magnets with a relatively weak uniaxial anisotropy might lie in the region of the phase diagram exhibiting three successive transitions, the one indicated, {\it e.g.\/},  by a dashed line in Fig.1. Hence, we consider  in the next section the effects of applied magnetic fields for a system in this regime.

\begin{figure}[ht]
\begin{center}
\includegraphics[scale=0.38]{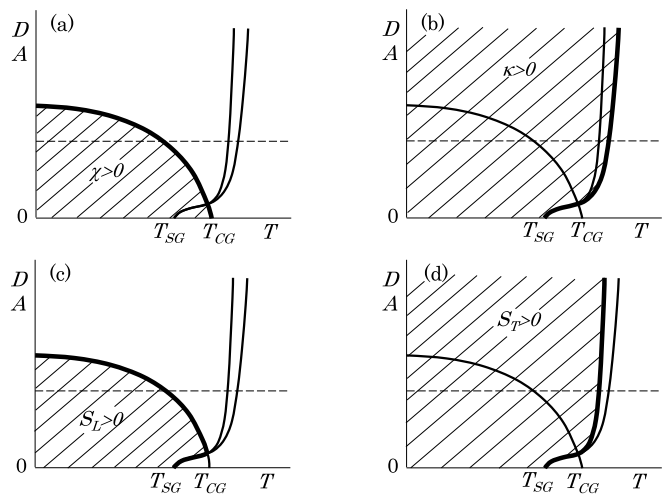}
\end{center}
\caption{
The region in the phase diagram of Fig.1 of the {\it XY\/}-like SG where each order parameter takes a nonzero value, the scalar chirality (a), the vector chirality (b), the transverse ($xy$) spin component (c), and the longitudinal ($z$) spin component (d).
}
\end{figure}

\section{Finite-field properties}

In this section, we analyze the effects of applied magnetic fields for a system with a relatively weak easy-plane-type uniaxial magnetic anisotropy, as indicated by the dashed line in Fig.1. We deal with two typical directions of applied magnetic fields, {\it i.e.\/}, the field applied along the hard axis ($\hat z$) and the one applied in the easy plane (say, along $\hat x$).

\subsection{Longitudinal fields}

 We begin with the longitudinal field $\vec H\parallel \hat z$. Applied longitudinal fields reduce the global symmetry of the system from the zero-field ones (I)$\sim $ (V) given in the previous section to, (ii) RT$_T$: SO(2)  `spin-rotation' around the magnetic-field axis ($\hat z$), and (iv) RF$_T$: $Z_2$ `spin-reflection' with respect to an arbitrary plane containing the magnetic-field axis. Other global Hamiltonian symmetries are suppressed under longitudinal fields.

 Let us analyze the fate of each zero-field transition. First, the vector-CG transition is expected to survive the longitudinal field as long as the field is not too strong, since the chiral $Z_2$ symmetry associated with RF$_T$ is retained even under longitudinal fields (RT$_L^{(\pi)}$ is suppressed due to the loss of RF$_L$). One also expects that replica symmetry is kept under weak magnetic fields, since it is unrelated to any global symmetry. In zero field, the transverse SG order was associated with `spin-rotation' symmetry RT$_T$. Even under longitudinal fields, RT$_T$ is retained so that the transverse SG transition still persists as a thermodynamic transition. Finally, the longitudinal SG transition, which was associated with `spin-reflection' symmetry RF$_L$ in zero field, would be gone under longitudinal fields, since the relevant symmetry is lost. For weaker fields, however, a rather rapid growth of the longitudinal component might still be visible as a crossover line.

\begin{figure}[ht]
\begin{center}
\includegraphics[scale=0.3]{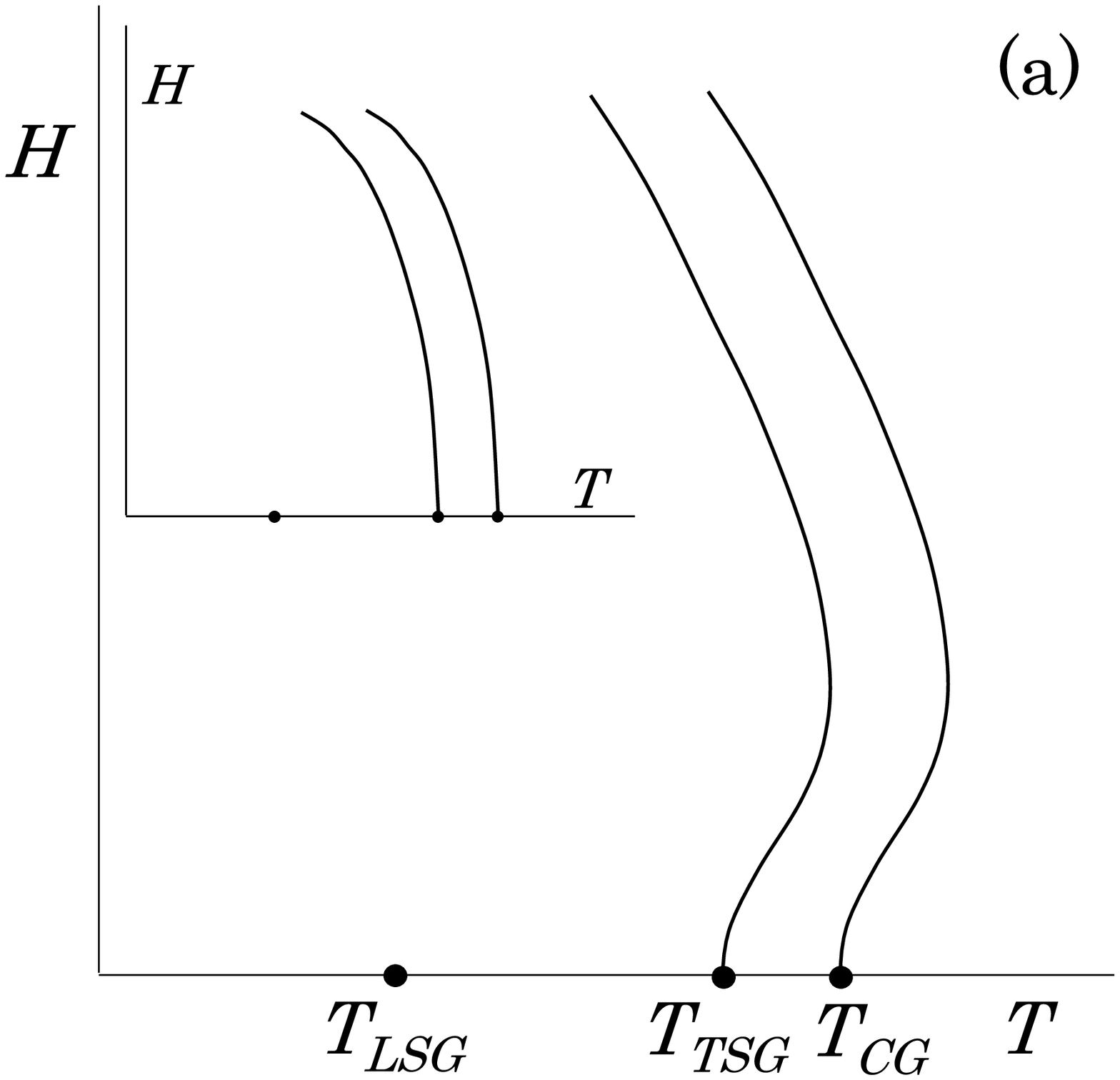}
\includegraphics[scale=0.3]{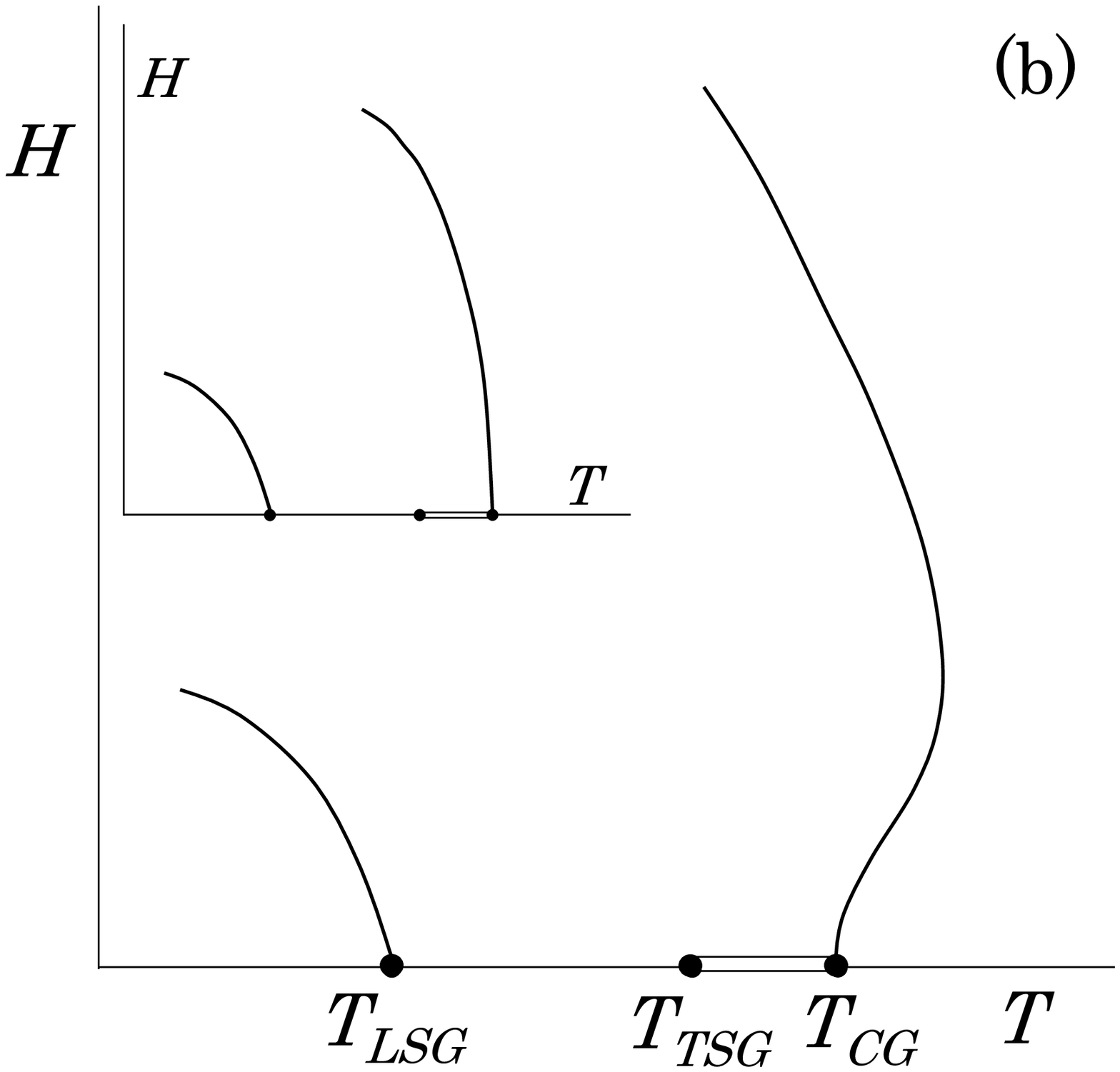}
\end{center}
\caption{
 Magnetic phase diagram of the {\it XY\/}-like SG with an easy-plane-type uniaxial magnetic anisotropy in the magnetic field versus the temperature plane. The uniaxial magnetic anisotropy is assumed to be moderately weak corresponding to the dashed line in the phase diagram of Fig.1. Magnetic field is applied either along the hard axis ($\hat z$) (a), or within the easy-plane ($\hat x$) (b).  $T_{CG}$, $T_{TSG}$ and $T_{LSG}$ represent the chiral-glass, the transverse spin-glass and the longitudinal spin-glass transition temperatures, respectively. Main panel and inset illustrate different forms of the upper phase boundary: See the text for details. The double bond between $T_{CG}$ and $T_{TSG}$ represent the vector-CG state where the spin and the chirality are decoupled.
}
\end{figure}

 On the basis of these symmetry observations, we construct a possible magnetic phase diagram in the magnetic field versus the temperature plane as in Fig.3(a). Concerning the form of each phase boundary at low fields, we follow the argument of Refs.\cite{Kawamura10}. Since the symmetries broken at the upper CG transition are common between in zero and in nonzero fields, {\it i.e.\/}, global $Z_2$ chiral symmetry associated with RF$_T$ and replica symmetry RS, the transition line $T_{CG}(H)$ is expected to exhibit a regular quadratic dependence on the field intensity as $T_{CG}(H)\approx T_{CG}(0) + c H^2 + \cdots$.

 The coefficient of the quadratic term $c$ could be either positive or negative on general grounds. When an easy-plane-type uniaxial anisotropy is sufficiently weak, a weak longitudinal field tends to suppress out-of-plane spin fluctuations and enhances  the spin-component perpendicular to the applied field, which might lead to  a positive $c$.  When an easy-plane-type uniaxial anisotropy gets stronger, on the other hand, spins tend to be contained in an $xy$-plane already in zero field by the uniaxial anisotropy, with negligible out-of-plane spin fluctuations. In such a case, an applied longitudinal field would just pull the spin along the field direction, which might lead to a negative $c$. Both forms of the CG phase boundary are indicated in Fig.3(a), $c>0$ in the main panel and $c<0$ in the inset. Hence, experimental system would exhibit a phase diagram of either type, depending on the strength of the easy-plane-type anisotropy.

 Concerning  the lower transverse SG transition line, the symmetry broken there is also common between in zero and in nonzero fields, {\it i.e.\/}, global RT$_T$. Hence, the transition line $T_{TSG}(H)$ at weak fields is also expected to exhibit a regular quadratic dependence on the field intensity. Similar arguments as made above can also be made concerning the coefficient of the quadratic term.

\subsection{Transverse fields}

 Next, we move on to the analysis of transverse magnetic fields $\vec H\parallel \hat x$, assuming again that an easy-plane-type uniaxial anisotropy is moderate. In this case, global Hamiltonian symmetries  are, (iii') RT$_L^{(\pi)}$: $Z_2$ `spin-rotation' of $\pi$ around the magnetic-field axis ($\hat x$), (iv') RF$_T$: $Z_2$ `spin-reflection' with respect to a plane containing the hard axis and the magnetic-field axis ($zx$-plane), and (v) RF$_L$: $Z_2$ `spin-reflection' with respect to the easy plane ($xy$-plane). Note that, while $\pi$ `spin-rotation' RT$_L^{(\pi)}$ and  `spin-reflection'  RF$_T$ still persist, the `spin-rotation' axis and the `spin-reflection' plane are now limited to the magnetic-field axis ($\hat x$) and to the $xz$-plane, respectively, due to the loss of RT$_T$.

 The fate of each zero-field transition under transverse fields is as follows. First, the vector-CG transition persists accompanied with an RSB, since the chiral $Z_2$ symmetry associated with RF$_T$ is retained even under transverse fields (RT$_L^{(\pi)}$ is also retained since RF$_L$ is retained). By contrast, at least in its simplest scenario, the transverse SG transition in zero field might be smeared out under transverse fields, since the relevant symmetry RT$_T$ is lost. In fact, the transverse spin component $S_T$ now orders already at the CG transition due to the loss of RT$_T$. In this sense, in contrast to the zero-field case, {\it the CG transition under transverse fields is simultaneously the transverse SG transition\/}. This is somewhat reminiscent of the ``spin-chirality recoupling'' phenomena observed in the Heisenberg SG with random magnetic anisotropy where the random magnetic anisotropy induces the simultaneous SG order at the scalar-CG transition \cite{Kawamura92,Kawamura10}. Here in the {\it XY\/} case, an applied transverse field induces the simultaneous transverse SG order already at the vector-CG transition.  Finally, the longitudinal SG transition, which was associated with `spin-reflection' symmetry RF$_L$ in zero field, persists even under transverse fields, since the relevant symmetry is retained. 

 On the basis of these observations, we construct a magnetic phase diagram in the magnetic field versus the temperature plane as in Fig.3(b). Since the symmetries broken at the upper CG transition are common between in zero field and in nonzero fields, the transition line $T_{CG}(H)$ is expected to exhibit a regular quadratic dependence on the field intensity $H$ as $T_{CG}(H)\approx T_{CG}(0) + c H^2 + \cdots$. Again, the coefficient $c$ could be either positive or negative. As argued above, $c$ is likely to be positive if an easy-plane-type uniaxial anisotropy is sufficiently weak, while it would be negative if the uniaxial anisotropy gets stronger.  The lower transition line associated with the longitudinal SG order is also expected to exhibit a regular quadratic dependence, since the  broken symmetry RF$_L$ is common between in zero field and in nonzero transverse fields.

\section{Effects of random magnetic anisotropy}

In this section, we analyze the effects of random magnetic anisotropy described by (3) and (4), which would inevitably exist in real materials to some extent. In most situations, however, its magnitude is weaker than that of the easy-plane-type uniaxial anisotropy.

In the presence of such random magnetic anisotropy, among global Hamiltonian symmetries, only time-reversal symmetry is kept in zero field, while all symmetries are lost under applied fields. By contrast, replica symmetry is likely to persist, since it is unrelated to any global Hamiltonian symmetry.

\subsection{Zero-field properties}

 First, we discuss  zero-field properties. As mentioned, in case of the isotropic Heisenberg SG, the random magnetic anisotropy is expected to cause ``spin-chirality recoupling'' phenomenon \cite{Kawamura92,Kawamura10}. Namely, the spin, once decoupled from the chirality in the fully isotropic case, is mixed into the chirality in the presence of weak random magnetic anisotropy. One might  see this from a symmetry consideration: The only global symmetry left in the presence of random magnetic anisotropy is $Z_2$ time-reversal symmetry, which is also the chiral symmetry flipping the sign of the scalar chirality. Hence,  even in the presence of weak random magnetic anisotropy, the CG transition essentially of the same nature as in the zero-field case accompanied by the $Z_2$ time-reversal symmetry plus RSB could still persist, while the loss of $SO(3)$ `spin-rotation' symmetry induces the simultaneous SG order already at the CG transition point, smearing out the SG transition of the isotropic system. In this recoupling phenomena, it is essential that the $Z_2$ chiral symmetry spontaneously broken at the CG transition is nothing but the $Z_2$ time-reversal symmetry, an only global symmetry retained in the presence of random magnetic anisotropy. 

 By contrast, in the strongly {\it XY\/} case with larger $D$ or $A$ value where the relevant chirality is the vector chirality rather than the scalar chirality, such a spin-chirality recoupling mechanism due to the random magnetic anisotropy does not operate. This is because the $Z_2$ symmetry spontaneously broken at the vector-CG transition is not a time-reversal symmetry, but is a reflection symmetry RF$_T$, which flips the sign of the vector chirality $\kappa_z$.   $Z_2$ time-reversal symmetry is broken not at the vector-CG transition $T_{CG}$, but at the SG transition  $T_{SG} (<T_{CG})$.  In the presence of random magnetic anisotropy, the $Z_2$ chiral symmetry associated with `spin-reflection' RF$_T$ is lost, though replica symmetry might still be preserved. Then, the vector-CG transition of the {\it XY\/} SG could still persist, but as a pure RSB transition changing its character from its isotropic counterpart. $Z_2$ time-reversal symmetry is eventually broken at the lower SG transition $T_{SG} (<T_{CG})$. Thus,  in the {\it XY\/} SG, there is no ``spin-chirality recoupling'' operative even in the presence of random magnetic anisotropy. The key difference from the Heisenberg case is that the chiral $Z_2$ symmetry of the {\it XY\/} SG is associated with the vector chirality rather than with the scalar chirality, and has nothing to do with $Z_2$ time-reversal symmetry.

 The above considerations suggest a phase diagram of the {\it XY\/}-like SG with random magnetic anisotropy in the uniaxial anisotropy versus the temperature plane as given in Fig.4. In the Heisenberg limit $D\rightarrow 0$ or $A\rightarrow 0$, there occurs a ``spin-chirality recoupling'' due to the random anisotropy. As a result, a single SG transition, which is essentially a disguised CG transition, takes place. In the {\it XY\/} regime beyond a multicritical point, the system exhibits successive phase transitions  with decreasing the temperature, the vector-CG transition at a higher temperature $T=T_{CG}$ and the SG transition associated primarily with the transverse spin component at a slightly lower temperature $T=T_{TSG}$. Note that a small amount of longitudinal spin component is already induced at $T=T_{TSG}$ due to the random magnetic anisotropy. Both transitions are genuine thermodynamic transitions, the CG transition being an RSB transition and the SG transition being a $Z_2$ time-reversal symmetry breaking transition. At a still lower temperature, a sharp crossover associated with rapid growth of the longitudinal spin component might occur as long as the random magnetic anisotropy is weak enough. This is a remnant of the longitudinal SG transition realized in a system without random magnetic anisotropy shown in Fig.3(b), and is indicated by the dotted curve in Fig.5(b).

\begin{figure}[ht]
\begin{center}
\includegraphics[scale=0.37]{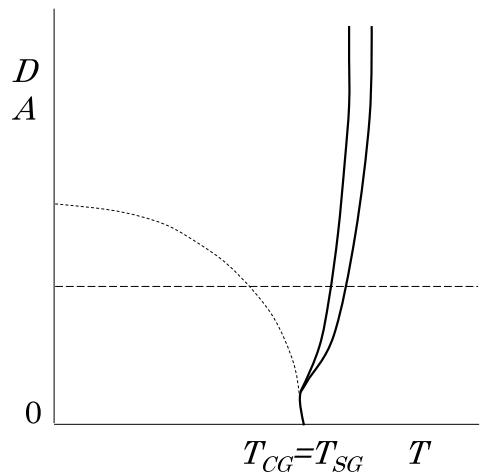}
\end{center}
\caption{
Phase diagram of the {\it XY\/}-like SG with a weak random magnetic anisotropy in the uniaxial anisotropy versus the temperature plane. $T_{CG}$ and $T_{SG}$ represent the chiral-glass and the spin-glass transition temperatures of the Heisenberg system.  
}
\end{figure}

 Hence, with decreasing the temperature in zero field, a system indicated by the dashed line in Fig.4 to which typical {\it XY\/}-like magnets like CdMn \cite{Murayama} and Mn$_x$Ni$_{1-x}$TiO$_3$ \cite{Ito,Kimura} correspond to, exhibits first the vector-CG transition at $T_{CG}$ at which the spin still remains paramagnetic, then the SG transition at $T_{TSG}$ associated primarily with the transverse spin component (but with a small amount of longitudinal spin component induced by the random magnetic anisotropy), and finally a sharp crossover at a lower temperature where the longitudinal spin component grows rapidly.

\begin{figure}[ht]
\begin{center}
\includegraphics[scale=0.3]{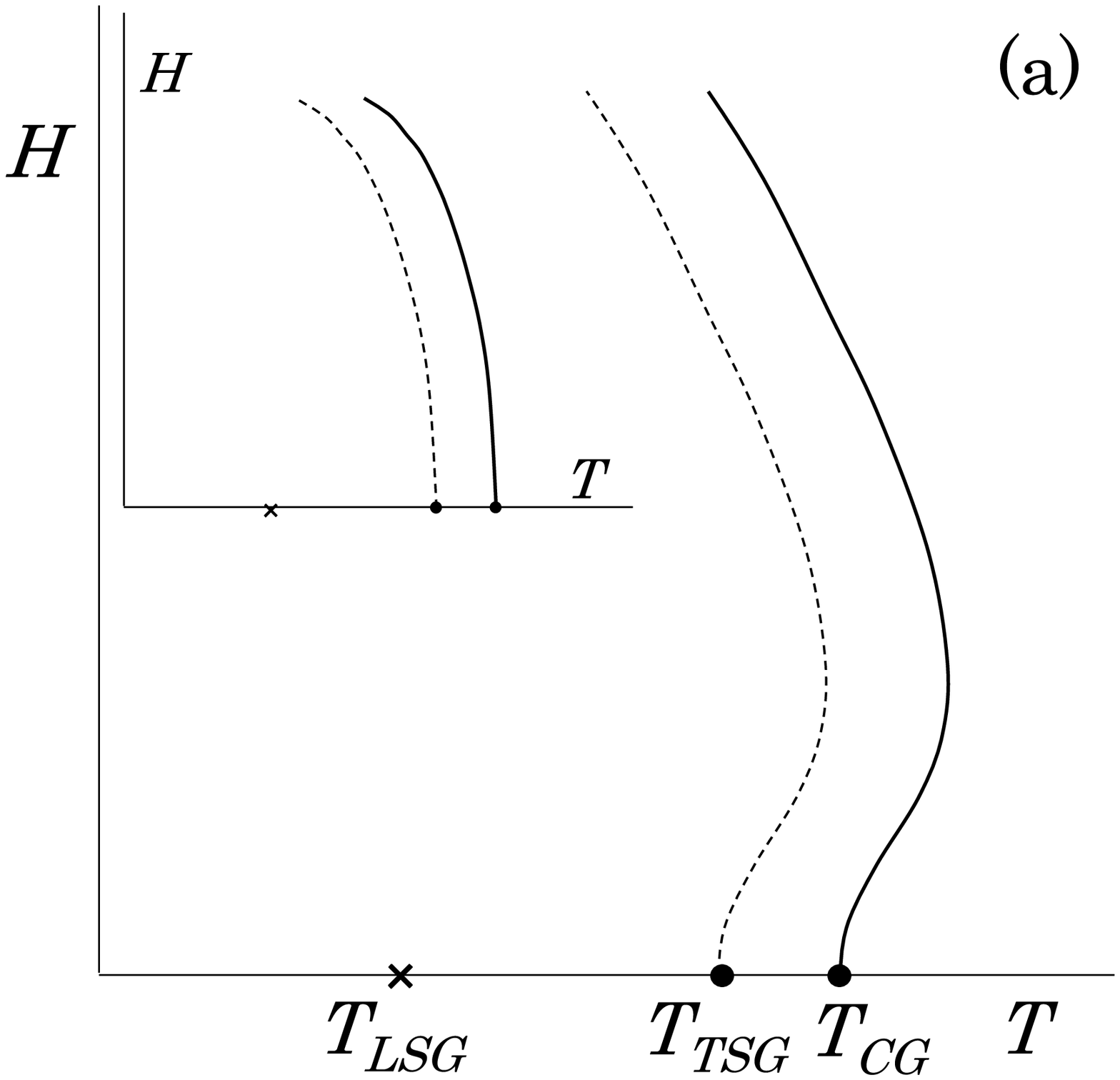}
\includegraphics[scale=0.3]{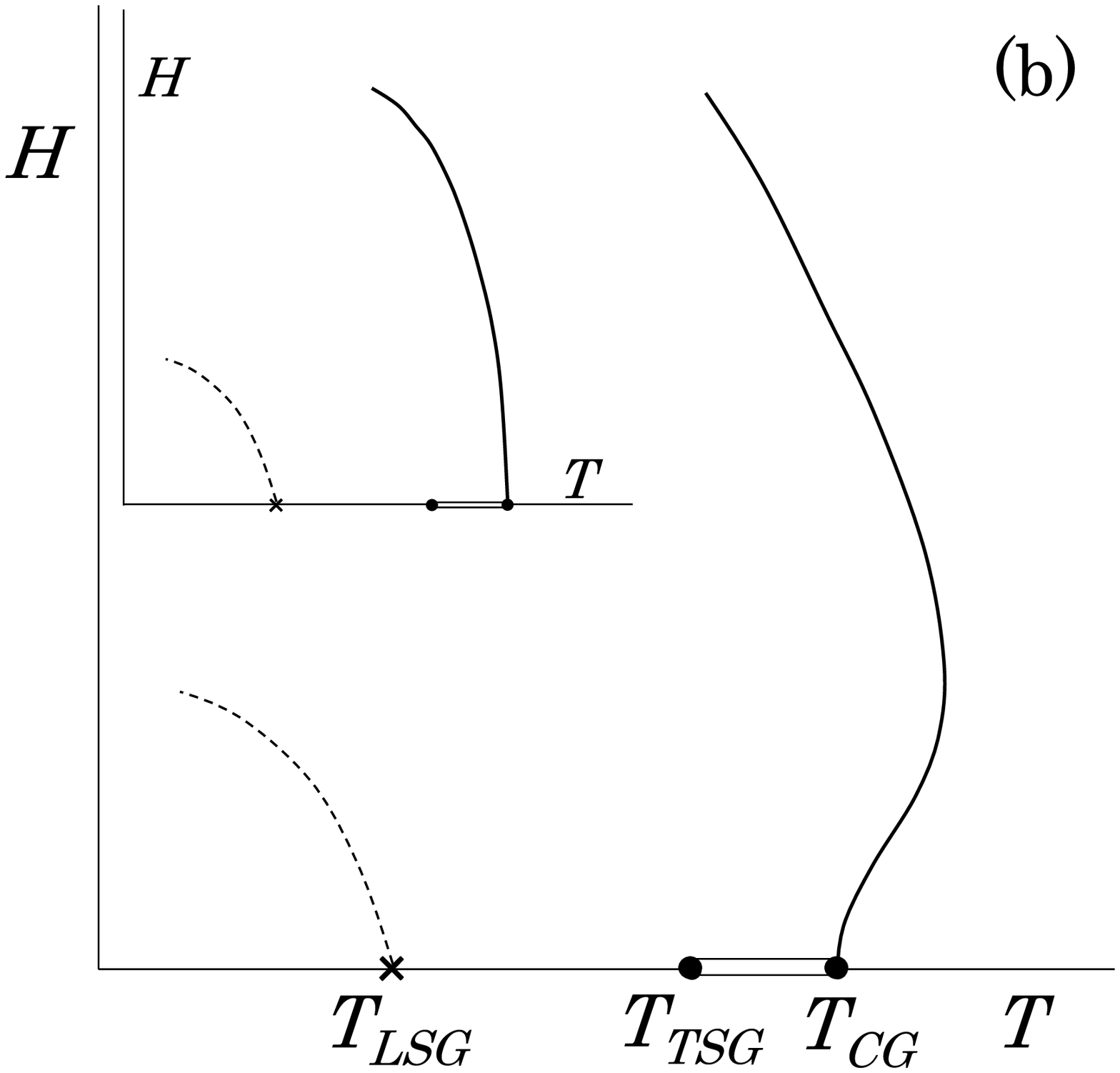}
\end{center}
\caption{
Magnetic phase diagram of the {\it XY\/}-like SG with a random magnetic anisotropy in the magnetic field versus the temperature plane. The uniaxial magnetic anisotropy is assumed to be moderately weak corresponding to the dashed line in the phase diagram of Fig.4. Magnetic field is applied either along the hard axis ($\hat z$) (a), or within the easy-plane ($\hat x$) (b).  $T_{CG}$, $T_{TSG}$ and $T_{LSG}$ represent the chiral-glass, the transverse spin-glass and the longitudinal spin-glass transition temperatures, respectively. Main panel and inset illustrate different forms of the upper phase boundary: See the text for details.  The double bond between $T_{CG}$ and $T_{TSG}$ represent the vector-CG state where the spin and the chirality are decoupled.
}
\end{figure}

\subsection{Finite-field properties}

 Next we analyze the effects of applied magnetic fields for a system with weak random anisotropy, which lies in the region of the phase diagram indicated by the dashed line in Fig.4. We analyze how the weak random magnetic modifies the magnetic field versus the temperature phase diagram of Fig.3. In the presence of both random magnetic anisotropy and applied magnetic fields, all global Hamiltonian symmetries  are gone, the only remaining symmetry being a replica symmetry. Hence, only true thermodynamic transition expected to remain under applied fields is a vector-CG transition, shown by the bold line in Fig.5. All other transition lines of Fig.3(a) and (b) turn into crossover lines. Such crossover lines are indicated by dotted curves in Fig.5(a) and (b). When the random magnetic anisotropy is sufficiently weak, which might be the case in most of real materials, a fairly sharp change could still occur there. Note that, in a system with random magnetic anisotropy in applied fields, a small amount of SG order, not only longitudinal but also transverse, should be induced already in the paramagnetic phase, reflecting the loss of any global Hamiltonian symmetry.

 Then, typical behaviors expected for a realistic {\it XY\/}-like SG material under applied fields would be as follows: For the longitudinal field applied along the hard axis ($\hat z$), the system exhibits with decreasing the temperature first an RSB vector-CG transition. The RSB nature of the transition would manifest itself most clearly as an onset of irreversibility of the chirality-related quantities. With further decreasing the temperature, the system exhibits a sharp crossover where the transverse SG order is sharply enhanced. Finally, the system exhibits the second crossover where the longitudinal SG order is further enhanced. 

 Similar behavior is expected also for the case of transverse fields applied within the easy plane ($xy$). Namely, with decreasing the temperature,  the system exhibits an RSB vector-CG transition accompanied with a sharp rise of the transverse SG order. Another crossover accompanied with a rise  of the longitudinal SG order might be visible at a lower temperature.

\section{Summary and discussion}

The ordering of {\it XY\/}-like SG magnets with an easy-plane magnetic anisotropy was studied based on  a symmetry consideration and recent results of numerical simulations on the pure Heisenberg \cite{VietKawamura09,Kawamura10} and {\it XY\/} \cite{KawaLiXY} SG models. The effects of an easy-plane-type uniaxial anisotropy, a random magnetic anisotropy and an applied magnetic field are analyzed, and magnetic phase diagrams are constructed. Various types of ordered phases, {\it e.g.\/},  the scalar-CG  phase, the vector-CG  phase, the coplanar SG phase, and the noncoplanar SG phase, are realized which are separated by thermodynamic  transition lines or crossover lines.

Interestingly,  in the {\it XY\/} regime in zero field, the spin-chirality decoupling persists even under the random magnetic anisotropy, escaping the spin-chirality recoupling mechanism which inevitably occurs in the Heisenberg SG. The vector-CG transition occurs at a higher temperature as an RSB transition and the transverse SG order occurs at a lower temperature associated with $Z_2$ time-reversal symmetry breaking, though the two temperatures might be rather close. This difference between the {\it XY\/} and Heisenberg regimes is originated from the difference in the nature of the vector and scalar chiralities, each being the order parameter of the CG transition of the {\it XY\/} and Heisenberg SGs. 

 In applied fields, all global Hamiltonian symmetries are gone in the presence of random magnetic anisotropy. As a result, small amount of SG order, both transverse and longitudinal, is induced even in the high-temperature paramagnetic phase. Only thermodynamic transition possible in applied fields is an RSB vector-CG transition. Sharp crossover could also occur in some cases.

 We note that the ordering behaviors similar to the ones observed here in the {\it XY\/}-like SG could also be expected even in the Ising-like SG,  {\it i.e.\/}, the vector SG with an easy-axis-type uniaxial magnetic anisotropy. The phase diagram of such an Ising-like SG is given in Fig.6 in the easy-axis-type uniaxial anisotropy versus the temperature plane. For moderate strength of the Ising-like anisotropy, the system exhibits with decreasing the temperature first the longitudinal SG order along the easy axis at which the symmetries RF$_L$ and RT$_L^{(\pi)}$ are spontaneously broken. Then, the transverse  components exhibit another transitions at lower temperature, which are very much similar to the ones realized in the {\it XY\/} SG. In particular, the vector-chirality and the transverse SG order might occur separately (the spin-chirality decoupling), each associated with the breaking of the symmetries RF$_T$ and RT$_T$, which coexists with the longitudinal SG order. In this way, essentially the same physics may be realized even in the ordering of Ising-like SG magnets.

\begin{figure}[ht]
\begin{center}
\includegraphics[scale=0.37]{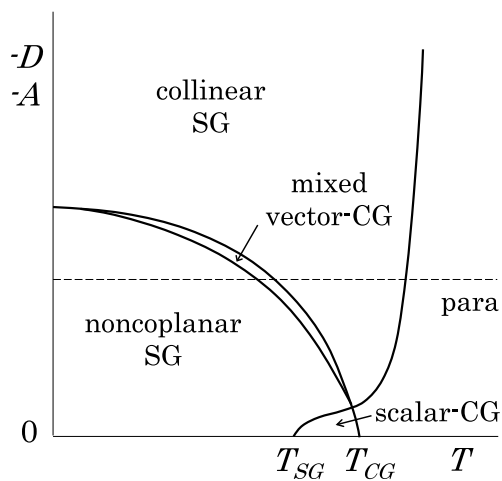}
\end{center}
\caption{
Phase diagram of the Ising-like SG with an easy-axis-type uniaxial magnetic anisotropy in the uniaxial anisotropy versus the temperature plane. $T_{CG}$ and $T_{SG}$ represent the chiral-glass and the spin-glass transition temperatures of the fully isotropic Heisenberg system.  
}
\end{figure}
 Finally, we wish to discuss experimental implications of our present results, particularly the way how to probe various transition or crossover lines.  First, we highlight the aforementioned spin-chirality decoupling phenomenon expected in the {\it XY\/} regime.  Since the order parameter at the CG transition is the vector chirality, and the spin is decoupled from the chirality even in the presence of random magnetic anisotropy, standard magnetic measurements are not suited to probe the CG transition (the TSG transition could be probed by the standard magnetic measurements). Recent studies have revealed that the electric polarization is often proportional to the vector chirality via an appropriate magneto-electric coupling (spin-current mechanism) \cite{Katsura}. Hence, electric polarization measurements might be utilized to detect the vector chirality ordering. Yet, since the vector chiral order here is a spatially random one, its detection might not be easy, at least in zero field. In applied fields, a uniform component of the vector chirality may be induced. Indeed, very recent measurements by Kimura and collaborators have revealed that, in an {\it XY\/}-like SG magnet Mn$_{1-x}$Ni$_x$TiO$_3$, applied magnetic fields induce  uniform electric polarization \cite{Kimura}.

 By contrast, the  SG transition or crossover lines, either transverse or longitudinal, might be detectable by standard magnetic measurements as an anomaly in the magnetization curve. The torque or the M\"ossbauer measurements might also be utilized to detect the transverse SG order. Hopefully, further experimental and theoretical studies will open up new horizon in the research of the chirality ordering in spin-glass magnets.

\medskip
 This study was supported by Grant-in-Aid for Scientific Research on Priority Areas ``Novel States of Matter Induced by Frustration'' (19052006). The author is thankful to T. Kimura, Y. Yamaguchi and D.X. Viet for useful discussion.

\end{document}